\documentclass[a4paper]{jpconf}
\usepackage{graphicx}
\usepackage{amsmath}
\begin{document}
\title{Machine-learning techniques applied to three-year exposure of ANAIS--112}

\author{I~Coarasa$^{1,2 }$, J~Apilluelo, J~Amar{\'e}$^{1,2 }$, S~Cebri{\'an}$^{1,2 }$, D~Cintas$^{1,2 }$, E~Garc\'{\i}a$^{1,2 }$, M~Mart\'{\i}nez$^{1,2,3 }$, M~A~Oliv{\'a}n$^{4 }$, Y~Ortigoza$^{1,2,5 }$, A~Ortiz~de~Sol{\'o}rzano$^{1,2 }$, J~Puimed{\'o}n$^{1,2 }$, A~Salinas$^{1,2 }$, M~L~Sarsa$^{1,2 }$ and P~Villar$^{1 }$}

\address{$^{1 }$ Centro de Astropart{\'{\i}}culas y F{\'{\i}}sica de Altas Energ{\'{\i}}as (CAPA), Universidad de Zaragoza, Pedro Cerbuna 12, 50009 Zaragoza, Spain}
\address{$^{2 }$ Laboratorio Subterr{\'a}neo de Canfranc, Paseo de los Ayerbe s.n., 22880 Canfranc Estaci{\'o}n, Huesca, Spain}
\address{$^{3 }$ Fundaci{\'o}n ARAID, Av. de Ranillas 1D, 50018 Zaragoza, Spain}
\address{$^{4 }$ Fundaci{\'o}n CIRCE, Av. de Ranillas 3D, 50018 Zaragoza, Spain}
\address{$^{5 }$ Escuela Universitaria Polit{\'e}cnica de La Almunia de Do\~{n}a Godina (EUPLA), Universidad de Zaragoza, Calle Mayor 5, La Almunia de Do\~{n}a Godina, 50100 Zaragoza, Spain}

\ead{icoarasa@unizar.es}

\begin{abstract}
ANAIS is a direct dark matter detection experiment aiming at the confirmation or refutation of the DAMA/LIBRA positive annual modulation signal in the low energy detection rate, using the same target and technique. ANAIS--112, located at the Canfranc Underground Laboratory in Spain, is operating an array of 3$\times$3 ultrapure NaI(Tl) crystals with a total mass of 112.5~kg since August 2017. The trigger rate in the region of interest (1-6 keV) is dominated by non-bulk scintillation events. In order to discriminate these noise events from bulk scintillation events, robust filtering protocols have been developed. Although this filtering procedure works very well above 2~keV, the measured rate from 1 to 2~keV is about 50\% higher than expected according to our background model, and we cannot discard non-bulk scintillation events as responsible of that excess. In order to improve the rejection of noise events, a Boosted Decision Tree has been developed and applied. With this new PMT-related noise rejection algorithm, the ANAIS--112 background between 1 and 2~keV is reduced by almost 30\%, leading to an increase in sensitivity to the annual modulation signal. The reanalysis of the three years of ANAIS--112 data with this technique is also presented.
\end{abstract}

\vspace{-0.7cm}
\section{Introduction}
The ANAIS (Annual modulation with NaI(Tl) Scintillators) experiment~\cite{anais2019Perf,anais2021Results} is intended to search for dark matter annual modulation with ultrapure NaI(Tl) scintillators at the Canfranc Underground Laboratory (LSC) in Spain, in order to provide a model independent confirmation or refutation of the signal reported by the DAMA/LIBRA collaboration~\cite{dama2020} using the same target and technique. This signal, observed for about 20 years with a statistical significance greater than 12$\sigma$, is in strong tension with the negative results of other very sensitive experiments~\cite{Schumann2019}, but a direct comparison using the same target material was still lacking.

ANAIS--112, consisting of nine 12.5~kg NaI(Tl) modules in a 3$\times$3 matrix configuration, accumulating 112.5~kg of mass in total, is taking data at LSC since August 2017. The ANAIS modules, built by Alpha Spectra Inc., stand out for incorporating a Mylar window in the lateral face allowing low energy calibration using external gamma sources, and for their exceptional optical quality, which added to the high efficiency Hamamatsu photomultipliers (PMTs) coupled to the crystals enable a light collection at the level of 15~photoelectrons/keV~\cite{anais2017LY} in all the modules. $^{109}$Cd sources are used every two weeks to calibrate the experiment and correct possible gain drifts. Low energy calibration is performed by combining the information from the $^{109}$Cd lines and from known lines present in the background at 3.20 and 0.87~keV (from $^{40}$K and $^{22}$Na crystal contamination, respectively). These events are tagged by coincidences with high energy depositions (1460.8 and 1274.5~keV, respectively) in a second module. The use of $^{22}$Na/$^{40}$K lines increases the reliability of the energy calibration in the region of interest (ROI, 1-6 keV), since they are actually either in the ROI or very close to the energy threshold, set at 1~keV~\cite{anais2019Perf}.

The trigger rate in the ROI is dominated by non-bulk scintillation events. Because of that, the development of robust protocols for the selection of events corresponding to bulk scintillation in sodium iodide is mandatory. They are based on the following criteria~\cite{anais2019Perf}: (1)~a pulse shape cut combining the fraction of the pulse area in [100,600]~ns after the event trigger and the logarithm of the amplitude-weighted mean time of the individual photoelectrons arrival times in the digitized window; (2)~an asymmetry cut by imposing more than 4~peaks in each PMT signal to remove events having a strongly asymmetric light sharing between the 2~PMTs; (3)~single-hit events; (4)~events arriving more than 1~s after the last muon veto trigger. Although this filtering procedure works very well above 2~keV, the measured rate from 1 to 2~keV is about 50\% higher than expected according to our background model~\cite{anais2019Bkg}, and we cannot discard non-bulk events as responsible of that excess. In order to improve the rejection of noise events between 1 and 2~keV, we have implemented a machine-learning technique based on a Boosted Decision Tree.

\section{Boosted Decision Tree}
Boosted Decision Tree (BDT) is a multivariate analysis technique which allows us to combine several weak discriminating variables into a single powerful discriminator~\cite{Coadou2010}. Decision trees have a binary structure with two classes: signal and noise\footnote{In general, background, but noise has been preferred to avoid confusion with radioactive background.}. During training, a sequence of binary splits using the discriminating variables is applied to the data in order to achieve the best separation between signal and noise. Depending on the output value of the decision tree, the event is classified as signal or noise. We use the purity of the terminal leaf as a separation criterion. The purity is defined as $p=\frac{s}{s+n}$, where $s$ ($n$) is the sum of weights of signal (noise) events that ended up in this leaf during training. If $p>0.5$, +1 is attached to the leaf and is classified as signal, otherwise the leaf is set to $-$1 and labelled as noise. The response of the decision tree with respect to fluctuations in the training sample can be stabilised using a boosting algorithm. The boosting of a decision tree~\cite{Freud1996} extends the concept from one tree to several trees, forming a forest. The trees are derived from the same training ensemble by reweighting events, and are finally combined into a single classifier which is given by a weighted average of the individual decision trees. In particular, we use AdaBoost (adaptive boosting) to implement the boosting algorithm.

Following the strategy of ANAIS--112 protocols, we separate the training into two phases:
\begin{itemize}
  \item In the first phase, we try to identify the scintillation events considering input parameters only depending on the shape of the total pulse from the two PMTs. As training populations, $^{109}$Cd calibration events between 1 and 2~keV are used as signal sample, whereas events from the Blank module are used as noise sample. The Blank module is similar to the nine ANAIS--112 modules, but without NaI(Tl) crystal, and is contained within a specific lead shielding next to ANAIS--112. It is coupled to 2~PMTs identical to those used in ANAIS modules and is integrated in the ANAIS--112 DAQ system. Consequently, the different events recorded in the Blank module have their origin in the PMTs and are very useful as a training population for non-bulk scintillation events.
  
  \item In the second phase, we use as input parameters those that describe the asymmetry in the light sharing among both PMTs for low energy events to separate the signal from the noise. Again, $^{109}$Cd calibration events are used as signal sample, while the background data are used as noise sample for training the BDT because they are dominated by PMT noise-like events in the very low energy region.
\end{itemize}

\begin{figure}[htb!]\vspace{-0.2cm}
\begin{minipage}{0.47\textwidth}
\includegraphics[width=\textwidth]{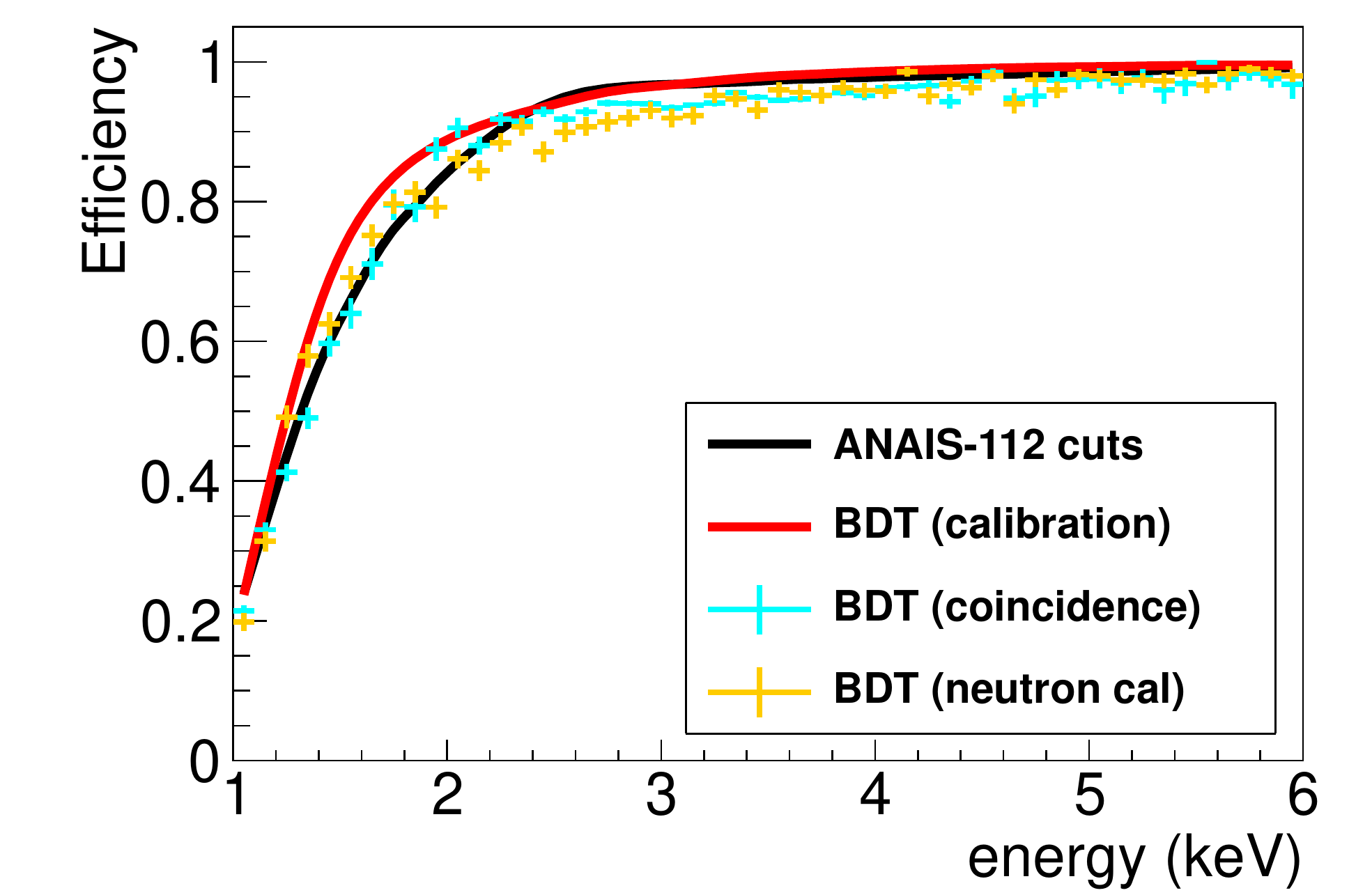}\vspace{-0.15cm}
\caption{\label{fig:eff}Average total efficiency obtained using BDT-BDT2 cut estimated from $^{109}$Cd calibration (in red), $^{22}$Na/$^{40}$K coincidences (in cyan) and neutron calibration (in orange). Efficiency with current ANAIS--112 protocols is shown in black.}
\end{minipage}\hspace{0.05\textwidth}%
\begin{minipage}{0.47\textwidth}
\includegraphics[width=\textwidth]{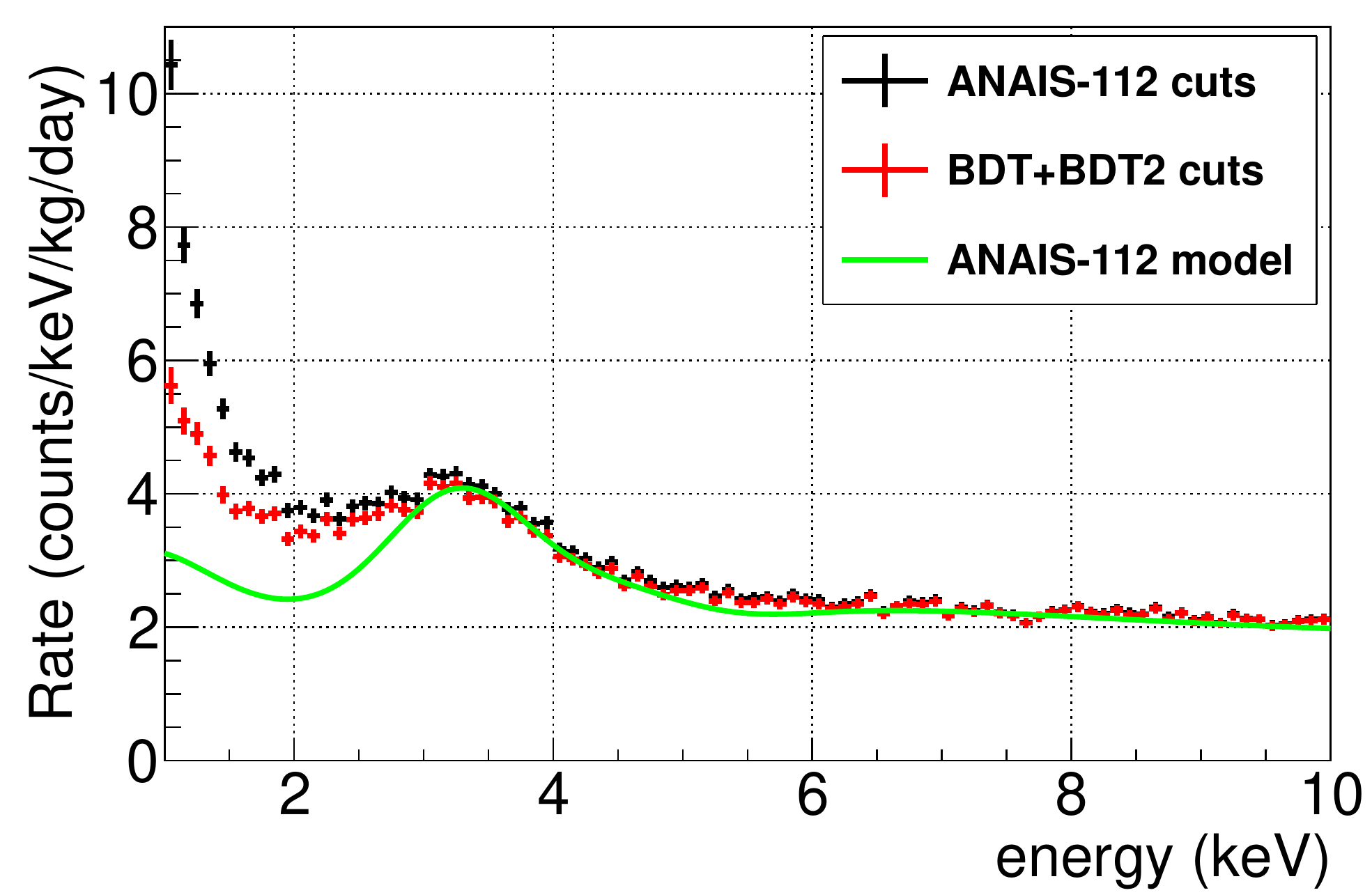}\vspace{-0.15cm}
\caption{\label{fig:LEspc}Total anticoincidence energy spectrum measured in the ROI after BDT-BDT2 event selection and efficiency correction (in red), using current ANAIS--112 filtering procedure (in black), and background model prediction (in green). Data correspond to the $\sim$10\% unblinded for the first year.}
\end{minipage}\vspace{-0.3cm}
\end{figure}

As a result of the training, we obtain two new variables (BDT and BDT2) that allow us to better discriminate noise from scintillation events. Straight line cuts on BDT and BDT2 parameters are used for the final event selection. This selection has been optimized to ensure the lowest total background in [1,2]~keV from the $\sim$10\% unblinded data of the first year of data taking, guaranteeing acceptance efficiencies equal to or greater than those obtained with current ANAIS--112 filtering procedures. We select BDT$>$0.10 and BDT2$>$0.10, and the corresponding efficiency is estimated for each detector independently by using $^{109}$Cd calibration single-hit events. The ratio of the events which pass the signal selection to the total events is the acceptance efficiency. The average efficiency of the BDT-BDT2 cut is shown in red in Figure~\ref{fig:eff}, while that obtained with the ANAIS--112 filtering protocols is displayed in black for comparison. Furthermore, we use two additional populations to cross-check the consistency of the selection criteria and efficiency estimation: the $^{22}$Na and $^{40}$K low energy events selected in coincidence with a high energy $\gamma$ (in cyan), and multiple-hit events from a recent neutron calibration run with $^{252}$Cf (in orange). There are small differences between them that are under study.

The total anticoincidence background spectrum in the ROI after event selection and efficiency correction is shown in red (black) for the BDT-BDT2 (previous ANAIS--112) filter in Figure~\ref{fig:LEspc}. We can observe that the BDT method developed significantly reduces the background level below 2~keV with respect to that obtained by the current ANAIS--112 procedure. In particular, the integral rate from 1 to 2~keV is 5.77$\pm$0.06 and 4.24$\pm$0.05~c/keV/kg/d for the ANAIS--112 filtering procedure and the BDT method, respectively, which represents a reduction of the background of around 25\%. Comparing this result with our background model (green line), we observe a 37\% discrepancy between 1 and 2~keV (compared to the 54\% observed previously), so there are still unexplained events below 2~keV, which could be related either with non-bulk scintillation events which have not been rejected by our BDT-BDT2 filtering procedure, or some background sources which have not been taken into consideration in our model. More work is underway in order to develop additional procedures for the rejection of the remaining PMT-noise events.

%efficiency and background

\section{Annual modulation analysis with BDT filtering}

\begin{figure}[htb!]
  \vspace{-0.3cm}
  \begin{minipage}{0.47\textwidth}
  \centering
  \includegraphics[width=\textwidth]{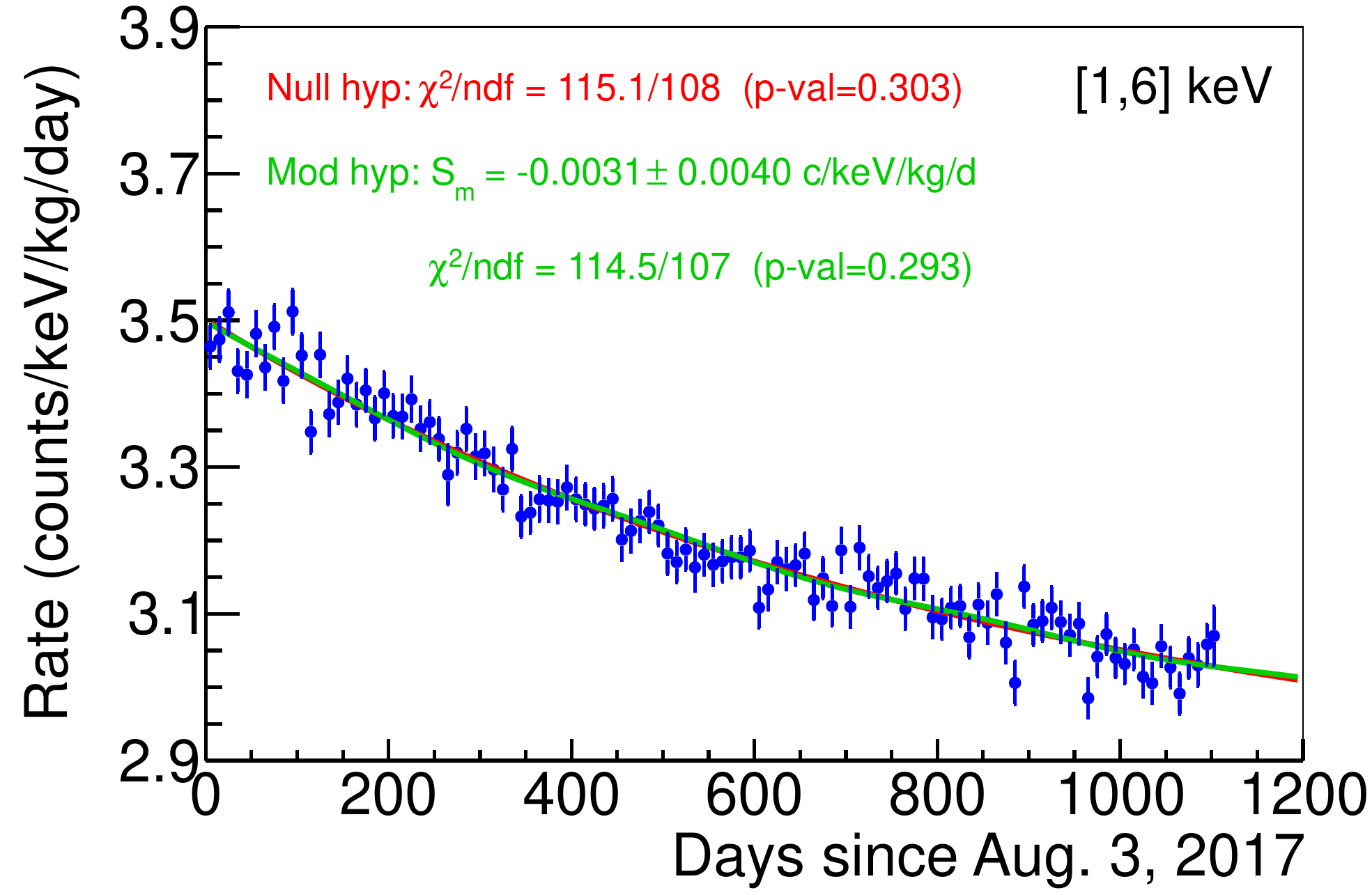}\\
  (a)
  \end{minipage}\hspace{0.05\textwidth}%
  \begin{minipage}{0.47\textwidth}
  \centering
  \includegraphics[width=\textwidth]{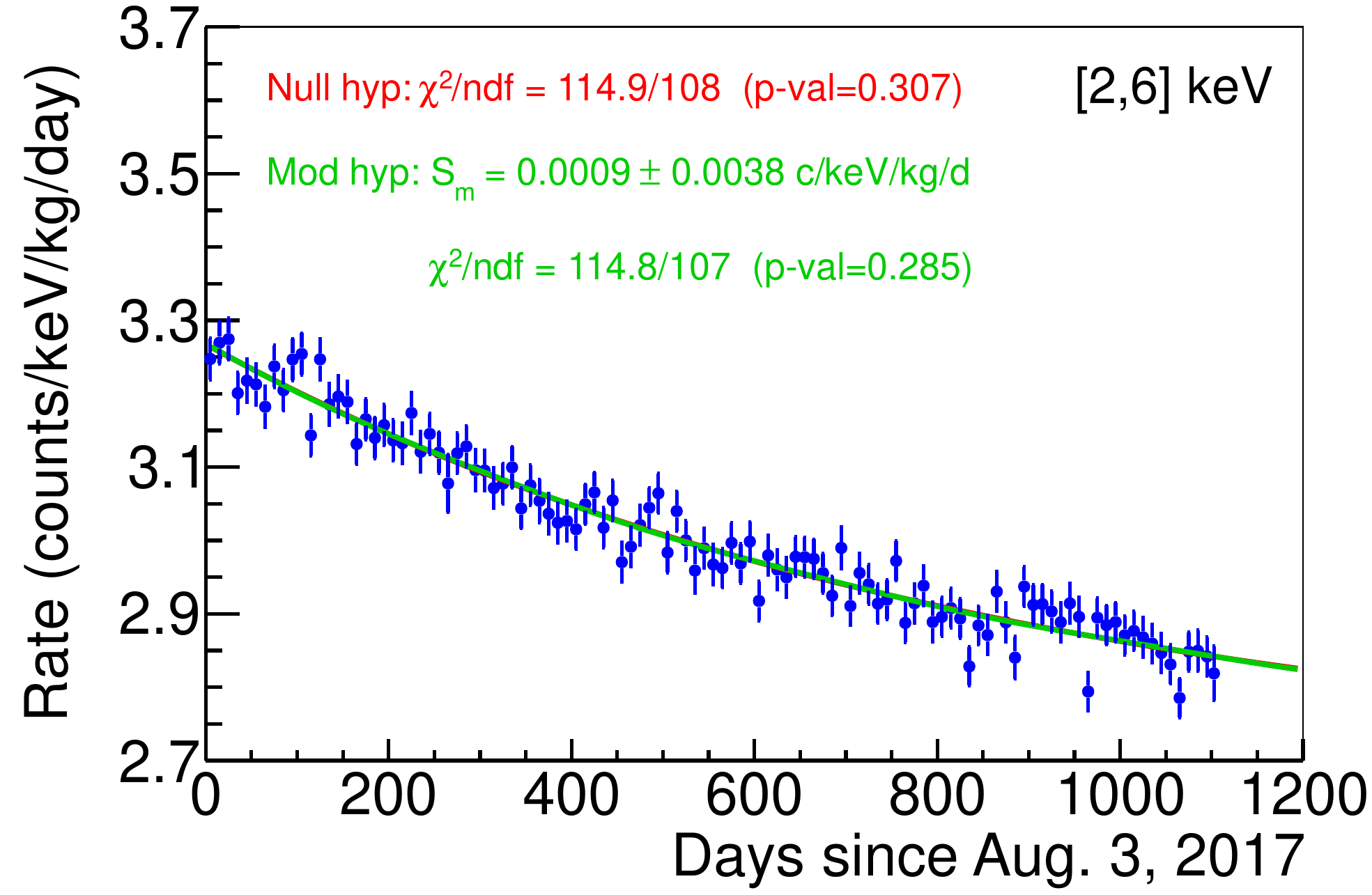}\\
  (b)
  \end{minipage}
  \caption{ANAIS--112 fit results for three years of data in [1,6]~keV (a) and [2,6]~keV (b) energy regions, both in the modulation (green) and null hypotheses (red) when the background is described by Equation~\ref{eq:rateEvolExpBDT}. Best fit $S_m$, $\chi^2$ and p-values are also shown.}
  \label{fig:rateEvolExpBDT}
\end{figure}

\begin{figure}
  \vspace{-0.4cm}
  \begin{minipage}{0.47\textwidth}
  \centering
  \includegraphics[width=\textwidth]{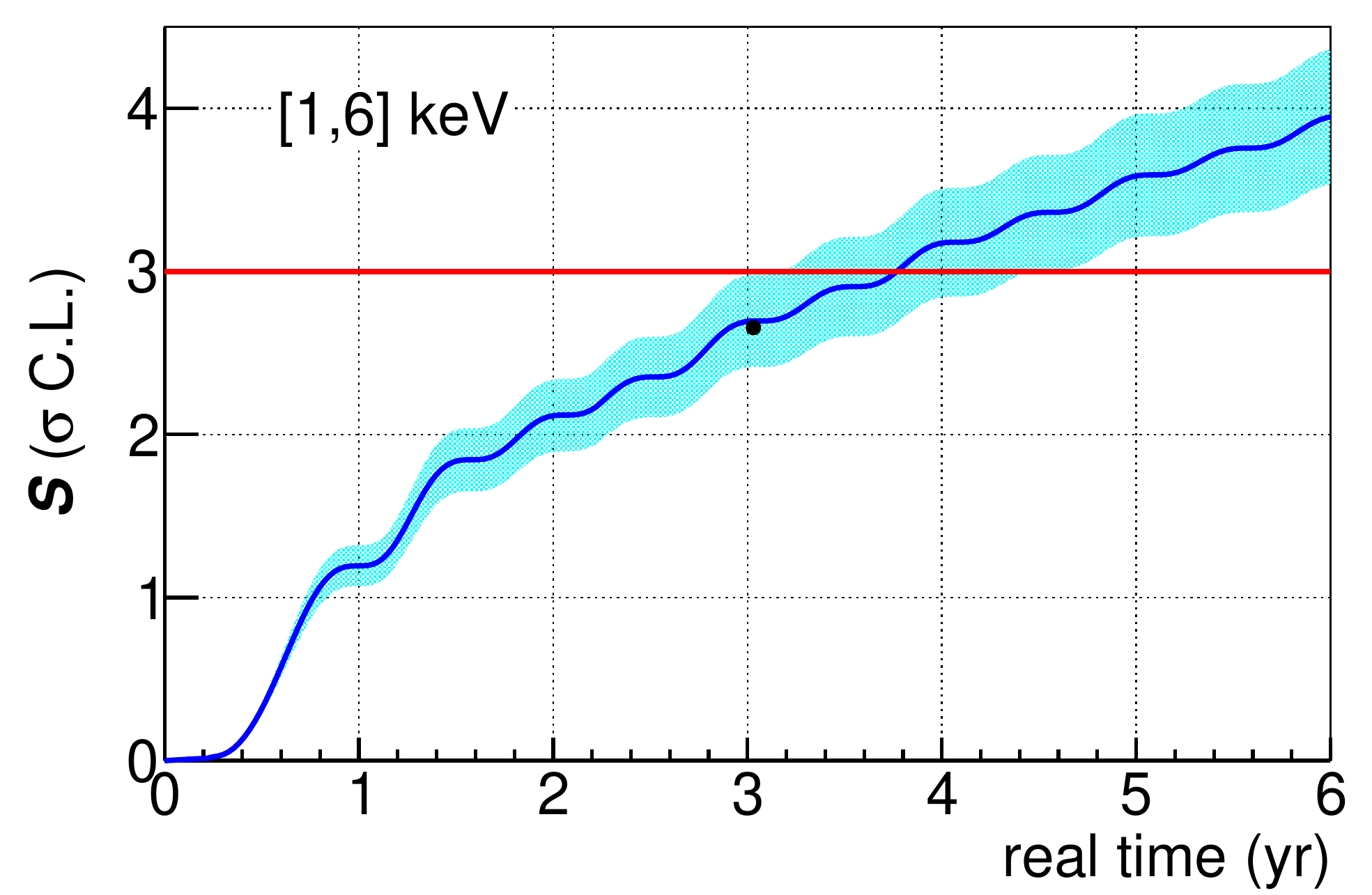}\\
  (a)
  \end{minipage}\hspace{0.05\textwidth}%
  \begin{minipage}{0.47\textwidth}
  \centering
  \includegraphics[width=\textwidth]{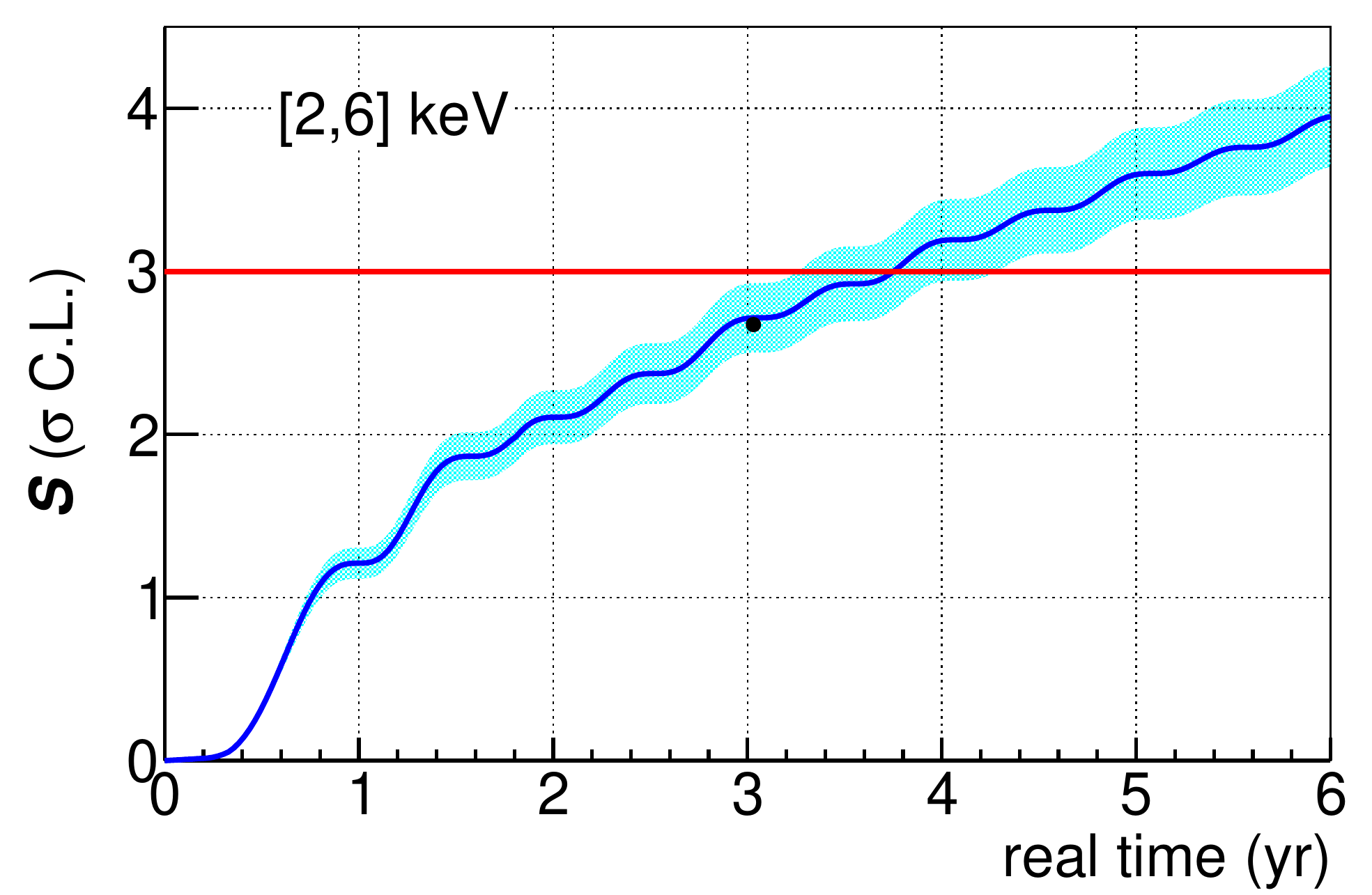}\\
  (b)
  \end{minipage}
  \caption{ANAIS--112 sensitivity to the DAMA/LIBRA signal recalculated using the background resulting after applying the BDT cut in $\sigma$ C.L. units as a function of real time in the [1,6] keV (a) and [2,6] keV (b) energy regions. The black dots are the sensitivities derived from the reanalysis of the three years of ANAIS--112 data using the BDT method. The cyan bands represent the 68\% C.L. DAMA/LIBRA uncertainty.}
  \label{fig:sensBDT_exp}
  \vspace{-0.4cm}
\end{figure}

Using the BDT filtering method, we have carried out the reanalysis of the three years of ANAIS--112 data searching for annual modulation in the same regions as DAMA/LIBRA has published ([1,6]~keV and [2,6]~keV), using an exposure of 322.83~kg$\times$yr. Data from all the modules are added together and grouped in 10-day time bins. Then, a least squares fit of the rate of events is performed, modelling the data as:
\begin{equation}
  \mathcal{R}(t_i) = R_0\cdot(1 + fe^{-t_i/\tau}) + S_m\cos\omega(t_i-t_0),
  \label{eq:rateEvolExpBDT}
\end{equation}
where $\mathcal{R}(t_i)$ is the expected rate of events in the time bin $t_i$; $R_0$, $f$ and $\tau$ are free parameters, while the modulation amplitude $S_m$ is fixed to 0 for the null hypothesis and left unconstrained (positive or negative) for the modulation hypothesis; $\omega=2\pi/365$~days$^{-1}$ and $t_0=-62.17$~days (corresponding the cosine maximum to 2$^{nd}$ June, when taking as time origin 3$^{rd}$ August). The results of the $\chi^2$ minimization following Equation~\ref{eq:rateEvolExpBDT} are shown in Figure~\ref{fig:rateEvolExpBDT} for [1,6]~keV~(a) and [2,6]~keV~(b) energy regions. The $\chi^2$ and p-values of the fit for the null (modulation) hypothesis are also shown in red (green), together with the best fit for $S_m$. The null hypothesis is well supported by the $\chi^2$ test in both energy regions, with $\chi^2$/NDF=115.1/108 (p-value=0.303) from 1 to 6~keV and $\chi^2$/NDF=114.9/108 (p-value=0.307) between 2 and 6~keV. The best fits for the modulation hypothesis are $S_m=-0.0031\pm0.0040$~c/keV/kg/d and $0.0009\pm0.0038$~c/keV/kg/d for [1,6] and [2,6]~keV, respectively, both compatible with zero at 1$\sigma$, and incompatible with DAMA/LIBRA result at 3.4$\sigma$ (2.4$\sigma$). As can be seen in Figure~\ref{fig:sensBDT_exp}, these results (black dots) also confirm our sensitivity prospects updated with respect to those previously calculated~\cite{anais2019Sens} considering the background level after filtering with the BDT. We quote our sensitivity to DAMA/LIBRA as the ratio of DAMA modulation result over the standard deviation on the modulation amplitude derived from ANAIS--112 data ($\mathcal{S}=S_m^{\text{DAMA}}/\sigma(S_m)$). At present, the achieved sensitivity is at 2.6$\sigma$ (2.7$\sigma$) for [1,6]~keV ([2,6]~keV) energy region. In particular, the BDT method improves the sensitivity between 1 and 6~keV by 9.1\%, and by 2.6\% from 2 to 6~keV, with respect to that obtained with the established ANAIS--112 filtering protocols. Moreover, the 3$\sigma$ sensitivity goal will be reached before the scheduled 5~years of data taking.

%rateEvol fitExp BDT (1x2)

%SensPlot exp BDT (1x2)

\vspace{-0.5cm}
\section{Conclusions}
A new PMT-related noise rejection algorithm based on the BDT technique has been developed for the ANAIS--112 experiment. With this filtering procedure, a background level reduction of around 25\% has been achieved between 1 and 2~keV, but there is still a 37\% discrepancy with our background model, which could be related with non-bulk scintillation events which have not been rejected by our BDT filtering protocol. More work is underway in order to develop additional procedures for the rejection of the remaining PMT-noise events. Furthermore, the reanalysis of the three years of ANAIS--112 data taking with this technique has been presented. We obtain for the best fit a modulation amplitude of $-0.0031\pm0.0040$~c/keV/kg/d ($0.0009\pm0.0038$~c/keV/kg/d) in the [1,6]~keV ([2,6]~keV) energy region, supporting the absence of modulation in our data, and being incompatible with the DAMA/LIBRA result at 3.4$\sigma$ (2.4$\sigma$), with a present sensitivity of 2.6$\sigma$ (2.7$\sigma$). According with our sensitivity estimates, which are confirmed with the latter results, the ANAIS--112 experiment will reach the 3$\sigma$ sensitivity goal before completing the scheduled 5~years of data taking, achieving 4$\sigma$ in 6~years of measurement.

\vspace{-0.1cm}
\ack
This work has been financially supported by the Spanish Ministerio de Economía y Competitividad and the European Regional Development Fund (MINECO-FEDER) under Grant No. FPA2017-83133-P; the Ministerio de Ciencia e Innovaci\'on -- Agencia Estatal de Investigaci\'on under Grant No. PID2019-104374GB-I00; the Consolider-Ingenio 2010 Programme under Grants No. MultiDark CSD2009-00064 and No. CPAN CSD2007-00042; the LSC Consortium; and the Gobierno de Arag\'on and the European Social Fund (Group in Nuclear and Astroparticle Physics and I.~Coarasa predoctoral grant). We thank the support of the Spanish Red Consolider MultiDark FPA2017-90566-REDC. The authors would like to acknowledge the use of Servicio General de Apoyo a la Investigaci\'on-SAI, Universidad de Zaragoza, and technical support from LSC and GIFNA staff.

\vspace{-0.1cm}
\section*{References}
\bibliographystyle{iopart-num}
\bibliography{proceedingsTAUP2021}

\end{document}